\documentstyle[12pt,epsf,epsfig,wrapfig]{article}
\begin{document}
\hfill{RUB-TII-08/07}
\begin{center}
{\bfseries GAUGE INVARIANCE AND RENORMALIZATION-GROUP EFFECTS IN
           TRANSVERSE-MOMENTUM DEPENDENT PARTON DISTRIBUTION 
           FUNCTIONS\footnote{Invited talk presented by the first 
           author at the XII Advanced Research Workshop on High 
           Energy Spin Physics (DSPIN-07), September 3-7, 2007, 
           Dubna, Russia.}}

\vskip 5mm I.\ O.\ Cherednikov$^{1,2 \star}$,
           N.\ G.\ Stefanis$^{2 \dagger}$
\vskip 5mm
{\small
(1) {\it
Bogoliubov Laboratory of Theoretical Physics, JINR,
141980 Dubna, Russia
}\\
(2) {\it
Institut f\"{u}r Theoretische Physik II, Ruhr-Universit\"{a}t Bochum,
D-44780 Bochum, Germany
}\\
$\star$ {\it
E-mail: igor.cherednikov@tp2.ruhr-uni-bochum.de
}\\
$\dagger$ {\it
E-mail: stefanis@tp2.ruhr-uni-bochum.de
}
}
\end{center}

\vskip 5mm
\begin{abstract}
A range of issues pertaining to the use of Wilson lines
in integrated and transverse-momentum dependent (TMD) parton
distribution functions (PDF) is discussed.
The relation between gauge invariance and the renormalization
properties of the Wilson-line integrals is given particular attention.
Using an anomalous-dimensions based analysis in the light-cone gauge,
a generalized definition of the TMD PDFs is proposed, which employs a
cusped Wilson line, and contains an intrinsic ``Coulomb-like'' phase.
\end{abstract}

\vskip 8mm

\paragraph{Introduction.}
Various calculations in the last few years have addressed TMD PDFs, 
among others those in which a previously overlooked transverse gauge 
link was proposed \cite{JY02,BJY03,BMP03}. 
The sustained interest in integrated and unintegrated (TMD) PDFs lies
in the fact that they encapsulate the nonperturbative quark
dynamics of confinement and hence in their potential use in
phenomenological applications to be compared with experimental
data. 
But while integrated PDFs can be defined in a gauge-invariant way that 
is compatible with factorization theorems, the definition of TMD PDFs 
faces serious problems related to specific light-cone divergences (see, 
e.g., \cite{BR05,Col03}). 
These so-called rapidity divergences \cite{CS81} are related to 
lightlike Wilson lines (or the use of the light-cone gauge $A^+=0$) 
\cite{Col02,CRS07} and cannot be cured by ordinary ultraviolet (UV) 
renormalization alone. 
In addition, in order to recover the result found in the Feynman gauge, 
the advanced boundary condition has to be adopted to make the 
transverse gauge link reduce to unity \cite{BJY03}.

The basic statement of the presented work \cite{CS07} is this:
In order to define an unintegrated PDF that preserves gauge invariance
under the proviso of collinear factorization and multiplicative
renormalizability, we shift our attention from the Wilson lines to
their anomalous dimensions within the $\overline{\rm MS}$ scheme.
We will provide concrete arguments that the appropriate contour which
goes through light-cone infinity is a cusped one.
To compensate the associated anomalous dimension, we introduce into
the definition of the TMD PDF a soft counter term (in the sense of
Collins and Hautmann \cite{CH00,CM04,Hau07}) which generates the same
anomalous dimension but with opposite sign.
Hence, the total TMD PDF expression has the same one-loop anomalous
dimension as the one that would involve a straight lightlike
line between the quark operators.
Note, however, that such a gauge contour cannot be adopted because
the gluons originating from this would not be collinear with the
struck quark and hence they would cause a mismatch in the gluon
rapidities.

To substantiate our arguments, we write the standard expression
for the TMD PDF \cite{CS81} for a quark-to-quark distribution,
supplemented by a transverse gauge link \cite{BJY03}:
\begin{eqnarray}
    f_{q/q}(x, \mbox{\boldmath$k_\perp$})
& = &
   \frac{1}{2}
   \int \frac{d\xi^- d^2
   \xi_\perp}{2\pi (2\pi)^2}
   {\rm e}^{- i k^+ \xi^- + i {\bf k}_\perp \cdot{\bf \xi}_\perp}
   \left\langle
   q(p) |\bar \psi (\xi^-, \xi_\perp)
   [\xi^-, \mbox{\boldmath$\xi_\perp$};
   \infty^-, \mbox{\boldmath$\xi_\perp$}]^\dagger \right.
\nonumber\\
&& \times\left.[\infty^-, \mbox{\boldmath$\xi_\perp$};
   \infty^-, \mbox{\boldmath$\infty_\perp$}]^\dagger
     \gamma^+[\infty^-, \mbox{\boldmath$\infty_\perp$};
   \infty^-, \mbox{\boldmath$0_\perp$}]
   [\infty^-, \mbox{\boldmath$0_\perp$};0^-, \mbox{\boldmath$0_\perp$}]
   \right.
\nonumber\\
&& \times\left. \psi (0^-,\mbox{\boldmath$0_\perp$}) |q(p)
   \right\rangle \
   |_{\xi^+ =0}\  ,
\label{eq:tmd_definition}
\end{eqnarray}
where the gauge links are defined according to
\begin{equation}
  [ \infty^{-}, \mbox{\boldmath$z_{\perp}$}; z^{-},
  \mbox{\boldmath$z_{\perp}$}]
\equiv
  {\cal P} {\rm e}^{
                     i g \int_{0}^{\infty} d\tau \ n_{\mu}
                     \hat A^{\mu} (z + n \tau)
               }\, \ , \
[ \infty^{-}, \mbox{\boldmath$\infty_{\perp}$};
 \infty^{-}, \mbox{\boldmath$\xi_{\perp}$}]
 \equiv
 {\cal P} {\rm e}^{
                      i g \int_0^{\infty} d\tau \ l_{i}
                     \hat A_{i}
                     (\vec \xi_{\perp}
                     + l_{i} \tau)
               }
\label{eq:transverse-gauge-link}
\end{equation}
with analogous expressions for the other gauge links and where
$\mbox{\boldmath$l_i$}$ represents an arbitrary vector in the
transverse direction and ${\cal P}$ denotes path ordering.

Within the Collins-Soper approach \cite{CS81} ($n^2\neq 0$), the 
anomalous dimension of $f_{q/q}(x, k_\perp)$ is \cite{JMY04}
 \begin{equation}
    \gamma_{\rm CS}
 =
    \frac{1}{2} \
    \mu \frac{d}{d \mu} \
    \ln Z_f
    (\mu, \alpha_s; \epsilon) =
    \frac{3}{4} \frac{\alpha_s}{\pi} C_{\rm F} + O (\alpha_s^2)
=
    \gamma_{\rm smooth}\ ,
\label{eq:gamma-smooth}
\end{equation}
where $Z_f$ is the renormalization constant of
$f_{q/q}(x, k_\perp)$ in the $\overline{\rm MS}$ scheme.
Recall that all smooth contours off the light cone in the transverse
direction give rise to the same anomalous dimension due to the
endpoints of the so-called connector insertion \cite{Ste83}.

\begin{figure}
\centering
\includegraphics[scale=0.36,angle=90]{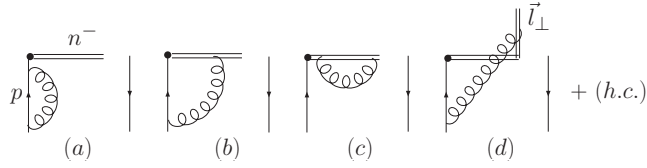}~~
\caption{One-loop gluon contributions to the UV-divergences of the
         TMD PDF.
         Double lines denote gauge links.
         Diagrams (b) and (c) are absent in the light-cone gauge.
\label{fig:se_gluon}}
\end{figure}

Figure \ref{fig:se_gluon} shows the one-loop diagrams, contributing to
$f_{q/q}(x, \mbox{\boldmath$k_\perp$})$
in the light-cone (LC) gauge
$(A \cdot n^-) = 0, \ {(n^-)}^2 = 0$.
The poles $1/q^+$ of the gluon propagator
\begin{equation}
   D_{\mu\nu}^{\rm LC} (q)
=
   \frac{1}{q^2} \Big( g_{\mu\nu}
  -\frac{q_\mu n^-_\nu + q_\nu n^-_\mu}{[q^+]}\Big) \ ,
\end{equation}
are regularized by
$1/[q^+]= 1/(q^+ \pm i \Delta)$, where $\Delta$ is small but finite.
In addition to the standard UV renormalization terms, one has UV
divergent contributions from diagrams (a) and (d) stemming from the
$p^+$-dependent term in
\begin{eqnarray}
    \Sigma_{\rm LC}^{\rm UV} (\alpha_s, \epsilon) =
    \frac{\alpha_s}{\pi}C_{\rm F} 2  \ \left[ \frac{1}{\epsilon }
    \left( \frac{3}{4}
    + \ln \frac{\Delta}{p^+} \right) - \gamma_E + \ln 4\pi \right]\, .
\label{eq:gamma_1}
\end{eqnarray}
Noting that the contribution associated with the transverse gauge
link at infinity (diagram Fig.\ 1(d)) exactly cancels against the
term entailed by the adopted pole prescription in the gluon propagator,
we find for the corresponding anomalous dimension
\begin{equation}
  \gamma_{\rm LC}
=
  \frac{\alpha_s}{\pi}C_{\rm F}\Bigg( \frac{3}{4}
  + \ln \frac{\Delta}{p^+} \Bigg)
=
  \gamma_{\rm smooth} - \delta \gamma \ .
\label{eq:gamma_2}
\end{equation}
Here $\delta \gamma$ is the term induced by the additional divergence
that has to be compensated by a suitable redefinition of the TMD PDF.
It is important to realize that
$p^+ = (p \cdot n^-) \sim \cosh \chi$ defines an angle $\chi$ between
the direction of the quark momentum $p_\mu$ and the lightlike vector
$n^-$ with $\ln p^+ \to \chi , \ \chi \to \infty$.
Hence, the ``defect'' of the anomalous dimension, $\delta \gamma$, can
be identified with the well-known cusp anomalous dimension \cite{KR87}
\begin{equation}
   \gamma_{\rm cusp} (\alpha_s, \chi)
= \frac{\alpha_s}{\pi}C_{\rm F} \ (\chi \coth \chi - 1 )
\ , \
 \frac{d}{d \ln p^+} \ \delta \gamma
= \lim_{\chi \to \infty}
  \frac{d}{d \chi} \gamma_{\rm cusp} (\alpha_s, \chi)
= \frac{\alpha_s}{\pi}C_{\rm F} \ .
\end{equation}
 \begin{figure}
\centering
\includegraphics[scale=0.5,angle=90]{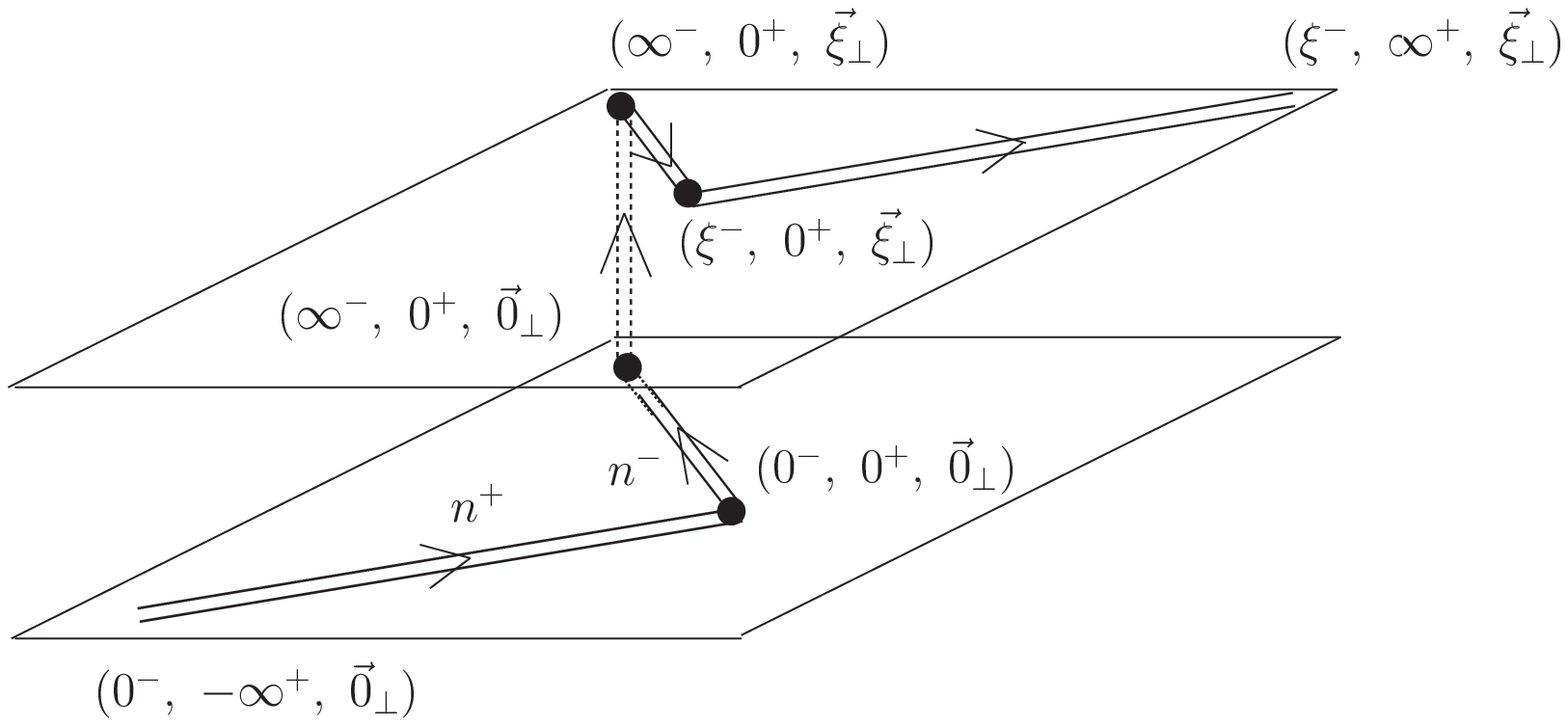}~~
\caption{The integration contour associated with the additional soft
         counter term.
\label{fig:contour}}
\end{figure}

Applying renormalization techniques for contour-dependent composite
operators \cite{Pol80,KR87,CD80} in order to treat angle-dependent
singularities, we introduce a compensatory soft term
\begin{equation}
  R
\equiv
  \Phi (p^+, n^- | 0) \Phi^\dagger (p^+, n^- | \xi) , \ \
  \Phi (p^+, n^- | \xi )
 =
  \left\langle 0
  \left| {\cal P} \exp\Big[ig \int_{\Gamma_{\rm cusp}}d\zeta^\mu
  \ t^a A^a_\mu (\xi + \zeta)\Big]
  \right|0
  \right\rangle
\label{eq:soft_definition}
\end{equation}
and evaluate it along the \emph{cusped} integration contour
$\Gamma_{\rm cusp}$, illustrated in Fig.\ \ref{fig:contour},
which is defined by ($n_\mu^-$ is the minus light-cone vector)
\begin{equation}
   \Gamma_{\rm cusp} : \ \ \zeta_\mu
=
  \{ [p_\mu^{+}s \ , \ - \infty < s < 0] \
   \cup \ [n_\mu^-  s' \ ,
  \ 0 < s' < \infty] \ \cup \
  [ \mbox{\boldmath$l_\perp$} \tau , \, \ 0 < \tau < \infty ] \}\, .
\label{eq:gpm}
\end{equation}
The one-loop gluon virtual corrections contributing to the UV 
divergences of $R$ are given by
\begin{equation}
  \Sigma_{R}^{\rm UV}
=
  - \frac{ \alpha_s}{\pi} C_{\rm F} \ 2 \left(  \frac{1}{\epsilon} \
  \ln \frac{\Delta}{p^+} - \gamma_E + \ln 4 \pi \right) \, .
\end{equation}
\label{eq:sigma_R}
This expression is equal, but with opposite sign, to the unwanted term
in the UV singularity, related to the cusped contour, calculated before.
This result enables us to redefine the conventional TMD PDF as follows:
\begin{eqnarray}
   f_{q/q}^{\rm mod}(x, \mbox{\boldmath$k_\perp$})
&&\!\!\!\!\!\!\! =
  \frac{1}{2}
   \int \frac{d\xi^- d^2
   \xi_\perp}{2\pi (2\pi)^2}
   {\rm e}^{- i k^+ \xi^- + i {\bf k}_\perp \cdot{\bf \xi}_\perp}
   \left\langle
   q(p) |\bar \psi (\xi^-, \xi_\perp)
   [\xi^-, \mbox{\boldmath$\xi_\perp$};
   \infty^-, \mbox{\boldmath$\xi_\perp$};]^\dagger
\right.\nonumber \\
&& \times   \left. [\infty^-, \mbox{\boldmath$\xi_\perp$};
   \infty^-, \mbox{\boldmath$\infty_\perp$};]^\dagger
   \gamma^+[\infty^-, \mbox{\boldmath$\infty_\perp$};
   \infty^-, \mbox{\boldmath$0_\perp$}]
   [\infty^-, \mbox{\boldmath$0_\perp$}; 0^-,\mbox{\boldmath$0_\perp$}]
\right.\nonumber \\
&& \times   \left.
   \psi (0^-,\mbox{\boldmath$0_\perp$}) |q(p)
   \right\rangle \
   \cdot
   \left[ \Phi(p^+, n^- | 0^-, \mbox{\boldmath$0_\perp$})
   \Phi^\dagger (p^+, n^- | \xi^-, \mbox{\boldmath$\xi_\perp$})
   \right] \, ,
\label{eq:tmd_re-definition}
\end{eqnarray}
The renormalization of
$
 f_{\rm ren}^{\rm mod}(x, \mbox{\boldmath$k_\perp$})
=
 Z_{f}^{\rm mod} (\alpha_s, \epsilon)
 f^{\rm mod} (x, \mbox{\boldmath$k_\perp$}, \epsilon)
$
yields the renormalization constant
\begin{eqnarray}
   Z_{f}^{\rm mod}
 =
  1 + \frac{ \alpha_s}{4\pi} C_{\rm F} \ \frac{2}{\epsilon}
  \left(- 3 - 4 \ln \frac{\Delta}{p^+}
  + 4 \ln \frac{\Delta}{p^+} \right)
 =
  1 - \frac{ 3\alpha_s}{4\pi}C_{\rm F} \ \frac{2}{\epsilon} \, .
\end{eqnarray}
which in turn provides the anomalous dimension
\begin{equation}
   \gamma_{f}^{\rm mod}
=
  \frac{1}{2} \mu \frac{d}{d\mu}
  \ln Z_{f}^{\rm mod} (\mu , \alpha_s, \epsilon)
= \frac{3}{4} \frac{\alpha_s}{\pi} C_{\rm F} + O(\alpha_s^2)
= \gamma_{\rm smooth} \, .
\end{equation}

To conclude, the soft counter term can be considered \cite{CS07} 
as that part of the TMD PDF which accumulates the residual effects of 
the primordial separation of two oppositely color-charged particles,
created at light-cone infinity and being unrelated to the existence of
external color sources, thus corresponding to an ``intrinsic Coulomb
phase'' that keeps track of the full gauge history of the colored
quarks \cite{JS90,Man62}.

\bigskip
\noindent
I.O.C. is supported by the Alexander von Humboldt Foundation.
This work was supported in part by the Deutsche Forschungsgemeinschaft
under grant 436 RUS 113/881/0, Russian Federation President's grant
1450-2003-2, and the Heisenberg--Landau Program 2007.

\end{document}